\documentclass[12pt]{iopart}

\usepackage{graphicx}
\def\etal{{\it et\thinspace al.}\ }
\def\eion{{(e~+~ion)}\ }

\def\fexvii{{\rm Fe~\sc xvii}\ }
\def\fexviii{{\rm Fe~\sc xviii}\ }
\def\fexix{{\rm Fe~\sc xix}\ }
\def\fexx{{\rm Fe~\sc xx}\ }
\def\fexxi{{\rm Fe~\sc xxi}\ }

\def\en{{$n$\ }}
\def\el{{$l$\ }}
\def\ii{{$i$\ }}
\def\dne{{$N_e$\ }}
\def\om{{$\omega$\ }}
\def\sig{{$\sigma$\ }}
\def\gam{{$\Gamma$\ }}

\newcommand{\be}{\begin{equation}}
\newcommand{\ee}{\end{equation}}
\begin{document}

\title[Photoionization and Opacity]{Photoionization and Opacity}

\author{A K Pradhan$^{1,2,3}$}

\address{$^1$ Department of Astronomy, $^2$ Chemical Physics Program, 
$^3$ Biophysics Graduate Program,
Ohio State University, Columbus, Ohio 43210, USA}
\vspace{10pt}
\begin{indented}
\item[]September 2022
\end{indented}

\begin{abstract}
Opacity determines radiation transport through material
media. In a plasma source the primary contributors to atomic opacity are
bound-bound line transitions and bound-free photoionization into the
continuum. We review the theoretical methodology for
state-of-the-art photoionization calculations based on the R-matrix
method as employed in the Opacity Project, the Iron Project, and
solution of the heretofore unsolved problem
of plasma broadening of autoionizing resonances
due to electron impact, Stark (electric
microfields), Doppler (thermal), and core-excitations.
R-matrix opacity calculations entail huge amount of atomic data and
calculations of unprecedented complexity.
It is shown that in high-energy-density (HED) plasmas
Photoionization cross sections become 3-D energy-temperature-density
dependent
owing to considerable attenuation of autoionizing resonance profiles.
Hence, differential oscillator strengths and
monochromatic opacities are redistributed in energy.
Consequently, Rosseland and Planck mean opacities are affected
significantly.
\end{abstract}

%
%
%
%
%

\section{Introduction}

Physically, the opacity depends on all possible intrinsic light-atom
interactions that may absorb, scatter, or re-emit photons emanating
from the source and received by the observer. In addition, the opacity
depends on external conditions in the source and the medium.
In recent years there have been a number of theoretical and experimental
studies of opacities (viz. \cite{m18,pain23,p18}).
Whereas photoionization and opacity are linked in all plasma sources,
we focus especially on
high-energy-density (HED) environments such as stellar interiors and
laboratory fusion devices, that 
are characterized by temperatures and
densities together, typically $T > 10^6K$ and densities $N > 10^{15}$
cm$^{-3}$. Computed atomic cross sections and transition proabilities are
markedly perturbed by plasma effects.

Monochromatic opacity consist of four terms
of bound-bound (bb), bound-free (bf), free-free (ff), and scattering
(sc):
\begin{equation}
 \kappa_{ijk}(\nu) = \sum_k A_k \sum_j F_j \sum_{i,i'}
[\kappa_{bb(}(i,i';\nu) +
\kappa_{bf}(i,\epsilon_{i'};\nu) + \kappa_{ff} (\epsilon_i, \epsilon'_{i'};
\nu) + \kappa_{sc} (\nu)]\ .
\label{eq:k}
\end{equation}
In Eq.~(1) $A_k$ is element abundance $k$, its ionization 
fraction $F_j$, $i$ and initial bound and final bound/continuum states
$i,i'$,  of a given atom; the $\epsilon$ represents electron energy in the 
continuum. To determine emergent radiation, a
harmonic mean $\kappa_R$, is defined,
{\it Rosseland Mean Opacity}
(RMO), with monochromatic opacity $\kappa_{ijk}(\nu)$ 

\begin{equation}
 \frac {1}{\kappa_R} = \frac{\int_0^\infty g(u) \kappa_\nu^{-1}
du}{\int_0^\infty g(u) du} \ \ \ \hbox{\rm with}\ \ \ g(u) = u^4 e^{-u} (1 - e^{-u})^{-2}.
\label{eq:RMO}
\end{equation}
Here, $g(u)$ is the derivative of the Planck function including
stimulated emission, 
$\kappa_{bb}(i,i') = (\pi e^2/m_ec) N_i f_{ii'} \phi_\nu$, and $\kappa_{bf} = N_i \sigma_\nu$. 
The $\kappa_\nu$ then depends on $bb$ oscillator strengths,
$bf$, photoionization cross sections $\sigma_\nu$, on the
equation-of-state (EOS) that gives level populations $N_i$. 
We describe large-scale 
computations using the coupled channel or close coupling
(hereafter CC) approximation implemented via the R-matrix (RM) method 
for opacity in 
Eq.~(\ref{eq:k}) primarily for: 
(i) the $bb$ transition probabilities and (ii) the $bf$ 
photoionization cross sections.

 In this review we focus on the $bf$-opacity,
and in particular on resonant phenomena
manifest in myriad series of autoionizing resonances that dominate
photoionization cross sections 
throughout the energy ranges of interest in practical applications.

\section{Photoionization}

  Photoionization (PI) of an ion $X^{+z}$ with ion charge $z$
into the \eion continuum is
\begin{equation}
 X^{+z} + h\nu \rightarrow X^{+z+1} +~e.
\end{equation}
PI also entails the indirect process
of resonances via formation of autoionizing (AI) doubly-excited states, and
subsequent decay into the continuum, as
\begin{equation}
h\nu + X^{+Z} \rightleftharpoons (X^{+Z})^{**} \rightleftharpoons 
X^{+Z+1} +~e 
\end{equation}
Infinite series of AI resonances are distributed throughout the 
photoionization cross
section and generally dominate at lower energies encompassing and
converging on to ionization
thresholds corresponding to excited levels of the residual ion in the
\eion continua.
A large number of photoionization cross section values for all bound levels 
are needed to compute plasma opacities.
Total photoionization cross section ($\sigma_{PI}$) of each bound 
level of the \eion system are required, from the ground state
as well as from all excited states.
Practically however we consider $n(SLJ) < $ 10, and
approximate relatively small number of energies below thresholds. 
Total $\sigma_{PI}$ corresponds to summed contribution 
of all ionization channels
leaving the residual ion in the ground and various excited states.

 AI resonances in photoionization cross sections
are dissolved by plasma density and temperature, 
resulting in an enhanced continuum background, as discussed later. 
However the strong and isolated 
resonances can be seen in absorption spectra. Moreover, a sub-class of
AI resonances corresponding to strong dipole transitions within the core
ion, known as Photoexcitation-of-core (PEC) or Seaton resonances,
correspond to the inverse process of dielectronic recombination
\cite{op,aas}.

Transition matrix for photoionization $S = <\Psi_F||{\bf D}||\Psi_{B}>$ 
is obtained from bound and continuum 
wave functions which give the line strength using the expression above.
Photoionization cross section is obtained as
\begin{equation}
\sigma_{PI} = {4\pi \over 3c}{1\over g_i}\omega S,
\end{equation}
where $\omega$ is the incident photon energy in Rydberg units.

\subsection{The Opacity Project and R-matrix Method}

Astrophysical opacity calculations using the RM method were 
intitated under the Opacity Project (circa 1983)
\cite{symp94,op,b11,aas}. 
The RM opacity codes were developed to compute large-scale 
and accurate bound-bound (bb) transition oscillator strengths,
and bound-free (bf) photoionization cross sections,
Considerable effort was devoted to precise delineation of the {\it
intrinsic} AI resonance profiles in terms of shapes, heights, 
energy ranges, and magnitudes 
determined by numerous coupled channels of the \eion system.

In the CC-RM method
the total \eion system is expressed 
in terms of the eigenfunctions of the 
target or core states and a free-electron
\begin{equation}
\Psi(E) = \mathcal{A} \sum_{i} \chi_{i}\theta_{i} + \sum_{j} c_{j} \Phi_{j}\ .
\label{eq:psi}
\end{equation}
The $\chi_{i}$ are target ion wavefunctions in a specific $S_{i}L_{i}$ 
state, $\theta_{i}$ is the free-electron wavefunction, and $\Phi_j$ are
bound channel correlation functions with coefficient
$c_j$ (viz. \cite{op,aas}). The coupled channel labeled as
$S_{i}L_{i}k_{i}^{2}\ell_{i}(SL\pi)$; $k_{i}^{2}$ is the
incident kinetic energy. In contrast, the distorted wave
approximation used in current opacity models neglects the summation over
channels in Eq.~\ref{eq:psi}, and therefore coupling effects
are not considered as in the RM method in an {\it ab inito} manner,
due to possibly hundreds to
thousands of coupled channels for complex ions. That approximation
in principle implies
neglect of quantum superposition in the distorted wave method, 
and interference that 
manifests in autoionizing resonance profiles.

The $bb$, $bf$ transition matrix elements for the \eion wave functions 
$\Psi_B(SL\pi;E)$ and $\Psi_F(SL\pi;E')$ respectively,
 bound state  $B$ and $B'$ line strengths (a.u.) are given by
\begin{equation}
{\protect S}(B;B') =
|\langle\Psi_B(E_B)||\mathbf{D}||\Psi_{B'}(E_{B'})\rangle|^2.
\end{equation}
For opacity computations we consider the $\mathbf{D}$  dipole operator,
since non-dipole transitions do not in general significant contributors. 
With the final continuum state represented by $\Psi_F(E')$ and the 
initial state by $\Psi_B(E)$, the photoionization cross section is 
\begin{equation}
\sigma_{\omega}(B;E') =
\frac{4}{3}\frac{\alpha\omega}{g_i}|\langle\Psi_B(E_B)||\mathbf{D}||\Psi_F(E')\rangle|^2.
\end{equation}
The $\omega$ is photon frequency and $E'$ is the photoelectron energy of the 
outgoing electron. The Breit-Pauli R-matrix (BPRM) incorporates 
relativistic effects using the
the Breit-Pauli (BP)
Hamiltonian for the \eion system in BPRM
codes in intermediate coupling with a pair-coupling 
scheme $S_iL_l(J_i)l_i(K_i)s_i (J\pi)$ \cite{rm1}, whereby
states $S_iL_i$ split into fine-structure levels $S_iL_iJ_i$.
Consequently, the number of 
channels becomes several times larger than the corresponding $LS$ coupling 
case. The IP work generally is based on BPRM codes, as for example the
large amount of radiative and collisional data in the database 
NORAD \cite{norad}. 

\subsection{R-Matrix Calculations for Opacities}

The $R$-Matrix codes employed in opacities calculations
are considerably different and extensions of the original $R$-Matrix codes 
\cite{b11,op,aas}. The OP codes were 
later extended under the Iron Project \cite{ip}
to incorporate relativistic effects and fine structure in the
Breit-Pauli approximation \cite{rm1}. 
The RM opacity
codes were further adapted with new extensions 
at Ohio State University for complete RM
opacity calculations \cite{np16,p18}. Fig.~1 shows the flowchart of 
the RM codes at the Ohio Supercomputer Center (OSC). The atomic
structure codes SUPERSTRUCTURE \cite{ejn} and CIV3 \cite{h75}, are first
utilized to obtain an accurate 
configuration-interaction representation of the core-ion states.
Next, The two $R$-Matrix codes STG1 and STG2 are 
employed to generate multipole integrals and algebraic coefficients 
for the \eion Hamiltonian corresponding to coupled 
integro-differential equations in the CC approximation. 
In the BPRM codes, the code RECUPD recouples the $LSJ$ pair coupling
representation incluing fine structure explicitly.
The total \eion Hamiltonian matrix is diagonalized in 
STGH. The $R$-Matrix basis functions and dipole 
matrix elements thus obtained are input to code STGB for bound state 
wavefunctions B, code STGF for continuum wavefunctions, $bb$ transitions
code STGBB, and code STGBF to compute photoionization cross sections. 
Code STGF(J) may also be 
used to obtain electron impact excitation collision strengths.

\begin{figure}
\includegraphics[height=5.5in,width=5.5in]{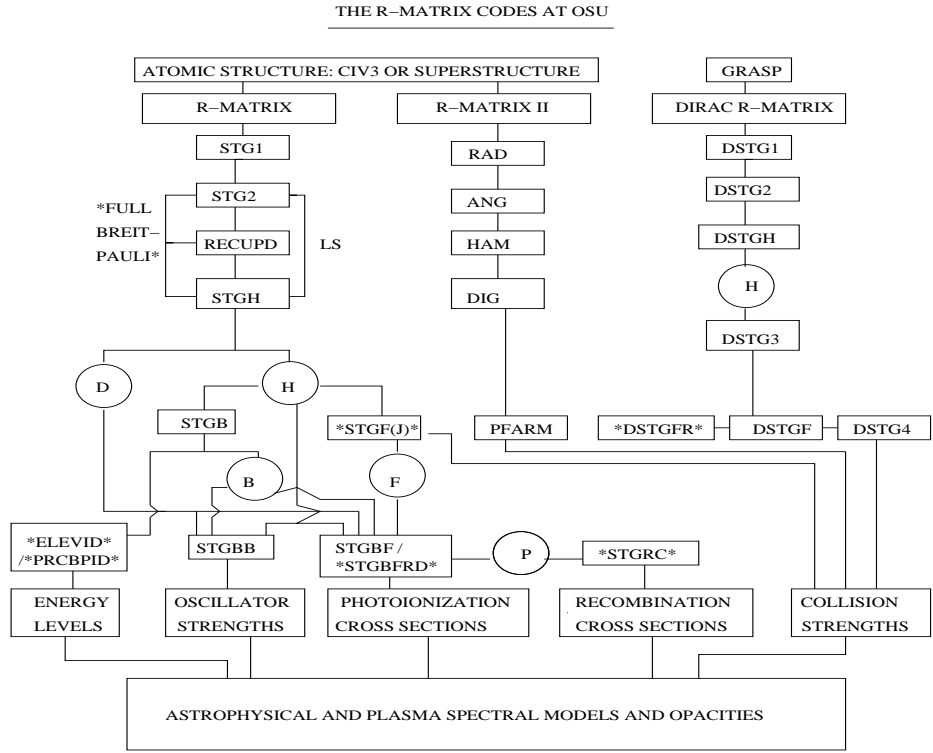}
\caption{The R-matrix codes for opacities calculations. Atomic data
produced is further processed by a suite of
equation-of-state, plasma broadening, and opacity codes to obtain 
monochromatic and mean opacities at each temperature and density
\cite{petal23}.
\label{fig:rm}}
\end{figure}

The immense complexity of RM calculations, compared to DW
method and atomic structure calculations, requires substantial 
computational effort and resources. In particular, inner-shell 
transitions are often dominant contributors to opacity. But those
could not be completed in OP work, except for outer-shell radiative
transtions using the RM or BPRM methods
due to computational constraints 
and then available high-performance computing platforms. Therefore, the
simpler DW method was used for most of the OP opacity calculations, 
such as in DW-type methods in other 
opacity models that also neglect channel couplings and hence {\it ab
initio} cconsideration of autoionizing resonances in the bound-free
continua. A prominent exemplar is
the extensive role of {\it photoexcitation-of-core} (PEC) 
resonances, or Seaton resonances \cite{op,aas}, associated with 
strong dipole transitions (viz. \cite{np16,p18} for Fe~XVII). 

 Despite unprecedented 
effort and advances, the OP-RM work faced several then
intractable difficulties that limited the scope of atomic calculations.
Primarily, the limitations were due to computational constraints which,
in turn, did not enable accounting for important physical effects and a 
complete
RM calculation of atomic opacities. The main features and
deficiencies of OP are as follows: (I) The calculations were in LS
coupling neglecting relativistic fine structure, (II) The close
coupling
wavefunction expansion for the target or the core ion in the \eion 
system included 
only a few ground configuration LS terms, (III) Inner-shell excitations
could not be included owing to the restricted target ion expansion that
precluded photoexcitation of levels from inner shells into myriad
resonances in the continua of the residual \eion system, (IV)
autoionizing resonances in bound-free photoionization cross sections were
delineated within the few excited target terms,
(V) Total angular and spin \eion
symmetries with large orbital angular-spin quantum numbers were not computed.
All of these factors are crucial for a
complete, converged and accurate opacity calculation.
As mentioned, the OP work initially began with the $R$-matrix codes,
albeit with very small wavefunction expansions \eion system,
usually limited to the ground configuration of the core ion. 
 Thus OP opacities incorporated a small subset of RM data.
Rather, most of the opacities contributions were obtained using atomic
structure codes and the Distorted Wave (hereafter DW)
approximation, similar to other opacity models [6-10]. 

 The first complete RM calculation leading up to the calculation
of opacities was carried out for the ion \fexvii that is of considerable
importance in determining the opacity at the base of
the solar convection zone (BCZ) (\cite{np16}, hereafter NP16). The solar
radius of the BCZ  has been
accurately determined through Helioseismology to be 0.713$\pm$0.001
R$_\odot$. 

Other new physical issues also emerged in RM calculations
for opacities. There are three major
problems that need to be solved: (A) convergence of large coupled channel
wavefunction expansions necessary to include sufficient atomic
structures manifest in opacity spectra, (B) completeness of high $n\ell$
contributions up to $n \equiv \infty$, and (C) attenuation of 
resonance profiles due to {\it intrinsic} autoionization broadening (included in
RM calculations in an ab initio manner) and {\it extrinsic} plasma
effects due to temperature and density, as generally considered for
bound-bound line opacity.
 
 RM photoionization calculations have been carried for several Fe ions
\cite{s18}. In particular, large-scale computations of cross sections and
transition probabilities have been done for Fe ions 
that determine iron opacity at the solar BCZ: \fexvii, \fexviii, \fexix,
\fexx and \fexxi (to be published; S.N. Nahar, private communication).

\subsection{R-matrix and Distorted Wave Methods}

Current opacity models employ the DW approximation or
variants thereof. 
 based on an atomic structure
calculation coupled to the continuum.
 Oscillator strengths and photoionization cross sections are
computed for all possible bound-bound and bound-free transitions 
among levels specified by electronic configurations included in the
atomic calculation. However, since the DW approximation includes only 
the coupling
between initial and final states, the complexity of interference
between the bound and
continuum wavefunction expansions involving other levels is neglected,
and so are the detailed profiles of autoionizing
resonances embedded in the continua. 
DW models employ
the independent resonance approximation that treats the bound-bound
transition probability independently from coupling to the continuum.
Apart from relative simplicity of atomic computations, 
the advantages of DW models is that
well-established plasma line broadening treatments may be used.

 On the other hand,
 RM opacities calculations are computationally laborious and 
time-consuming. However, as demonstrated in the erstwhile OP-RM work,
albeit severely limited in scope, coupling effects are important.
 Opacity in the bound-free continuum is dominated by autoionizing
resonances, as shown in recently completed works (viz.
\cite{np16,p18,p23}.
The most important consequence of neglecting detailed resonance profiles
in DW models and missing opacity is that {\it intrinsic} autoionizing
broadening and {\it extrinsic} plasma broadening thereof are not fully
accounted for. It has now been shown that AI resonances are broadened
much wider in the continuum than lines, and thereby enhance 
opacity significantly
\cite{np16,p18}.

 Recent work (\cite{d21}, D21) extended \fexvii RM
calculations by including more configurations than NP16a.
Whereas that confirmed
our earlier results for photoionization cross sections, 
D21 do not consider plasma broadening of autoionizing resonances
and therefore do not obtain a complete
description of bound-free opacity from RM calculations (discussed
below).
The unbroadened cross sections in D21 appear to
similar to ours but they did not compare those in detail with
previously published data in \cite{np16} for \fexvii, 
and publicly available from the electronic
database NORAD \cite{norad}.
Also, D21 report 10\% lower Rosseland mean opacities
than OP2005, which is at variance
with other DW models which are higher 
by up to a factor of about 1.5
(\cite{np16,p18}, possibly because of incomplete number of bound
\fexvii levels.

\section{Inner- and Outer-Shell Excitations}

Being simpler and based on pre-specified electronic
configurations as in atomic structure calculations, inner-shell
excitation DW data may be readily computed treating resonances as bound
levels in the continuum.
Although OP opacities were computed using DW
data, OP atomic codes were 
originally developed to implement the RM methodology that could not be
carried through owing to 
computational constraints. Most importantly it could not be employed for
opacities due to inner-shell excitations that are
dominant contributors because most electrons in complex ions 
are in closed shells and whose excitation energies lie above the first
ionization threshold, giving rise to series of autoionizing resoances,
and in particular PEC resonances due to 
strong dipole inner-shell trasitions
in the core ion \cite{np16,p23}. 
On the other hand, the much simpler DW treatment
in opacity models is readily implemented but is inaccurate in the
treatment of important resonance phenomena.
Extensive comparison of RM and DW calculations for \fexvii
considered herein, and implications for plasma opacities,
is given in \cite{np16,n11}.

\section{Plasma broadening of resonances}

Whereas line broadening has long been studied and its treatments are
generally and routinely incorporated in opacity models (viz. \cite{op}),
plasma broadening of autoionizing 
resonance profiles is not heretofore considered.
Attenuation of shape, height, energies, and magnitude of 
autoionizing resonances in photoionization cross
sections must be delineated in detail, as in the RM method, 
as function of density and temperature in order to determine
the distribution 
of total differential oscillator strength and structure of the
bound-free continua.

AI resonances are
fundamentally different from bound-bound lines as related to quasi-bound
levels with {\it intrinsic} quantum mechanical autoionization widths.
Broadening has significant contribution to mean opacities, enhancing the
Rosseland mean opacity by factors ranging from 1.5 to 3, as shown in
other works and discussed below \cite{p23}.
However, line broadening processes and formulae may be 
to develop a
theoretical treatment and computational algorithm outlined herein
(details to be presented elsewhere).
The convolved bound-free photoionization cross section of level \ii
may be written as:

\be \sigma_i(\omega) = \int \tilde{\sigma}(\omega') \phi
(\omega',\omega) d\omega', \ee

where \sig and $\tilde{\sigma}$ are the cross sections with
plasma-broadened
and unbroadened AI resonance structures, \om is the photon energy
(Rydberg atomic units are used throughout), and $\phi (\omega',\omega)$
 is the normalized 
Lorentzian profile factor in terms of the {\it total} width \gam due to
all
AI broadening processes included:

\be \phi (\omega',\omega) = \frac{\Gamma(\omega)/\pi}{x^2+\Gamma^2}, \ee

where $x \equiv \omega-\omega'$. The crucial difference with line
broadening is
that AI resonances in the \eion system correspond to and are due to
quantum mechanical interference between discretized continua
defined by excited core ion levels in a multitude of channels. The
RM method
(viz. \cite{b11,op,aas}), accounts for AI resonances in an \eion
system
with generally asymmetric profiles (unlike line profiles that are
usually symmetric).

Given $N$ core ion levels corresponding to resonance
structures,

\be \sigma(\omega) = \sum_i^N \left[ \int \tilde{\sigma}(\omega')
\left[ \frac{\Gamma_i(\omega)/\pi}{x^2 +
\Gamma_i^(\omega)}\right] d \omega' \right]
. \ee

 With $x \equiv \omega' - \omega $, the summation is over all excited
thresholds $E_i$ included in
the $N$-level RM wavefunction expansion, and corresponding
to total damping width $\Gamma_i$ due to all broadening processes.
The profile $\phi(\omega',\omega)$ is centered at each
continuum energy $\omega$, convolved over the variable $\omega'$ and
relative to each excited core ion threshold \ii.
In the present formulation we associate the energy to the effective
quantum number relative to each threshold $\omega' \rightarrow \nu_i$ to
write the total width as:

\begin{eqnarray}
\Gamma_i(\omega,\nu,T,N_e) &  = & \Gamma_c(i,\nu,\nu_c)+
\Gamma_s(\nu_i,\nu_s^*)\\
 & + &  \Gamma_d(A,\omega) + \Gamma_f(f-f;\nu_i,\nu_i'), \nonumber 
\end{eqnarray}

pertaining to
collisional $\Gamma_c$, Stark $\Gamma_s$, Doppler $\Gamma_d$, and
free-free transition $\Gamma_f$ widths
respectively, with additional parameters as defined below.
We assume a Lorentzian profile
factor that subsumes both collisional broadening due to electron impact,
and Stark broadening due to ion microfields, that dominate in
HED plasmas. This approximation should be valid since
collisional profile wings extend much wider as $x^{-2}$, compared to
the shorter range $exp(-x^2)$ for thermal Doppler, and $x^{-5/2}$ for
Stark
broadening (viz. \cite{op,p23}). In Eq. (11) the limits $\mp \infty$ are
then replaced by $\mp \Gamma_i/\sqrt{\delta}$; $\delta$ is
chosen to ensure the Lorentzian profile energy range
for accurate normalization.
Convolution by evaluation of Eqs. (1-3) is carried out for each
energy $\omega$ throughout the tabulated mesh of energies used to
delineate all AI resonance structures, for each cross section,
and each core ion threshold. We employ the following expressions for
computations:

\be \Gamma_c(i,\nu) \  = \ 5 \left( \frac{\pi}{kT} \right)^{1/2}
 a_o^3 N_e G(T,z,\nu_i) (\nu_i^4/z^2), \ee

where T, \dne, $z$, and $A$ are the temperature, electron density, ion
charge and atomic weight respectively, and $\nu_i$ corresponds to 
a given core ion threshold \ii: $\omega \equiv E =
E_i-\nu_i^2/z^2$ is a continuous variable. The Gaunt factor \cite{p23}
$G(T,z,\nu_i) = \sqrt 3/\pi [1/2+ln(\nu_i kT/z)]$
Another factor
$(n_x/n_g)^4$ is introduced for $\Gamma_c$
to allow for doubly excited AI levels with excited core
levels $n_x$ relative to the ground configuration $n_g$
(e.g. for \fexviii
$n_x=3,4$ relative to the ground configuration $n_g=2$).
A treatment of the Stark effect for complex
systems entails two approaches, one where both electron and ion
perturbations are combined, or separately (viz.
\cite{op,p23}) employed herein. Excited Rydberg levels are nearly
hydrogenic, the Stark effect is linear and
ion perturbations are the main broadening effect, though collisional
broadening competes increasingly with density
as $\nu_i^4$ (Eq.~13).
The total Stark width of a given
\en-complex is $\approx (3F/z)n^2$, where F is the plasma electric
microfields.
Assuming the dominant ion perturbers to be protons and density
equal to electrons, we take $F=[(4/3)
\pi a_o^3 N_e)]^{2/3}$, consistent with the Mihalas-Hummer-D\"{a}ppen
equation-of-state formulation \cite{op}.

\be \Gamma_s(\nu_i,\nu_s^*) =
[(4/3)\pi a_o^3 N_e]^{2/3} \nu_i^2. \ee

In employing Eq. (12) a Stark ionization parameter
$\nu_s^* = 1.2\times 10^3 N_e^{-2/15}z^{3/5}$ is introduced such
that AI resonances may be considered fully dissolved into the continuum
for $\nu_i > \nu_s^*$ (analogous 
to the Inglis-Teller series limit for plasma ionization of
bound levels). 
Calculations are carried out with and without
$\nu_s^*$ as shown later in Table~1. The Doppler width is:

\be \Gamma_d (A,T,\omega) = 4.2858 \times 10^{-7} \sqrt(T/A), \ee

where $\omega$ is {\em not} the usual line center but taken to be each
AI resonance energy. The last term $\Gamma_f$ in Eq. (5) accounts for
free-free
transitions among autoionizing levels with $\nu_i,\nu_i'$ such that 

\be X_i + e(E_i,\nu_i) \longrightarrow X_i' + e'(E_i',\nu_i'). \ee

The large number of free-free transition probabilities for $+ve$ energy
AI
levels $E_i,E_i' > 0$ may be computed using RM or atomic structure
codes (viz. \cite{z18}). 

 We utilize new results from an extensive
Breit-Pauli R-Matrix (BPRM) calculation
with 218 fine structure levels dominated by $n \leq 4$ levels of the core
ion \fexviii (to be reported elsewhere). A total of 587 \fexvii
bound levels ($E<0$) are considered, 
dominated by configurations $1s^22s^22p^6 (^1S_0),
1s^22s^p2p^qn\ell, [SLJ] \ (p,q
= 0-2, \ n \leq 10, \ \ell \leq 9, \ J \leq 12$). The core \fexvii
levels
included in the RM calculation for the (e~+~\fexviii) $\rightarrow
$\fexvii system
are:$1s^22s^22p^5 (^2P^o_{1/2,3/2}), 1s^22s^22p^q,n\ell, [S_iL_iJ_i] \
(p=4,5, \ n \leq 4, \ell \leq 3)$. The Rydberg series of
AI resonances correspond to $(S_iL_iJ_i) \ n \ell, \ n \leq 10, \ell
\leq 9$, with effective quantum number defined as a continuous variable
$\nu_i = z/\sqrt(E_i-E) \ (E>0)$, throughout the energy range up to the
highest
$218^{th}$ \fexviii core level; the $n=2,3,4$ core levels range from
E=0-90.7 Ry (\cite{np16}).
 The \fexvii BPRM calculations were carried out resolving the bound-free
cross sections at $\sim$40,000 energies for 454
bound levels with AI resonance structures.
Given 217 excited core levels of \fexviii, convolution
is carried out at each energy or approximately $10^9$ times for each
(T,$N_e$) pair.

\begin{figure}
\includegraphics[height=3.5in,width=5.5in]{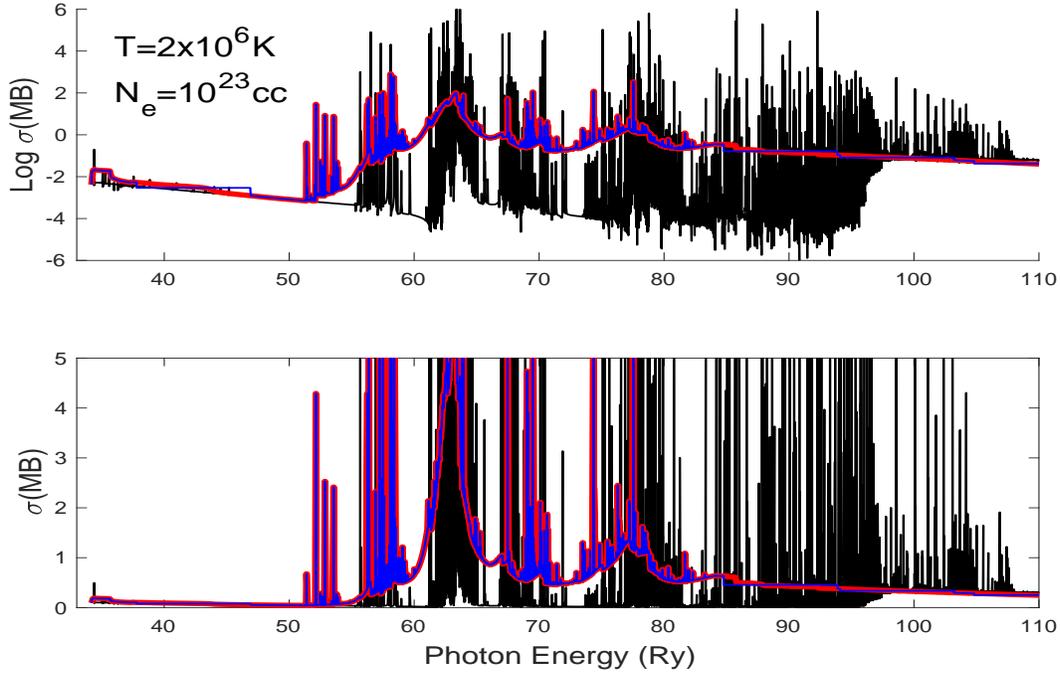}
\caption{Energy-temperature-density dependent photoionization cross
section of of highly excited bound level $2s^22p^53p \ ^2D_2$ of \fexvii
$\longrightarrow$ e \ + \fexviii,
due to plasma broadening of autoionizing resonances:
unbroadened --- black curve, broadened --- red and blue (see text). Top
panel: Log$\sigma$ (MB) in the full energy range up to the highest ionization
threshold of core ion \fexviii, bottom panel: Linear-scale $\sigma_{PI}$ 
in the energy range of the largest AI structures.
\label{fig:pbro}}
\end{figure}

Fig.~\ref{fig:pbro} displays detailed results for 
unbroadened photoionization cross section (black)
and plasma broadened (red
and blue, without and with Stark ionization cut-off)
The excited bound level
of \fexvii is $2s^22p^2 \ ^3D_2$ at temperature-density T=$2\times 10^6$K and
$N_e=10^{23}$cm$^{-3}$. The cross section is shown on the Log$_10$ scale in
the top panel, and on a linear scale in the bottom panel isolating the
energy region of highest and strongest AI resonances. 
The main features
evident in the figure are as follows.
(i) AI resonances show significant plasma broadening and smearing of
a multitude of overlapping Rydberg series at
The narrower high-\en \el resonances dissolve into
the continua but stronger low-\en \el resonance retain their asymmetric
shapes with attenuated heights and widths. (ii) At the
$N_e=10^{23}$cm$^{-3}$, close to that at the solar BCZ, 
resonance structures not only broaden but their strengths shift and 
are redistributed
over a wide range determined by total width
$\Gamma(\omega,\nu_i,T,N_e)$ at each energy $\hbar \omega$ (Eq. 12).
(iii) Stark ionization cut-off (blue curve) results in step-wise structures
that represent the average due to complete dissolution into continua.
(iv) Integrated AI resonance strengths are conserved, and are generally within
5-10\% of each other for all three curves in Fig.~\ref{fig:pbro},
It is found that the ratio of RMOs with and without plasma
broadening may be up to a factor of 1.6 or higher (\cite{p23}); recent
work for other ions shows the ratio may be up to factor of 3.

 The scale and magnitude of new opacity calculations is evident from the
fact that photoionization cross sections of 454 bound levels of \fexvii 
are explicitly calculated using the RM opacity codes,
1154 levels of \fexviii, and 899 levels \fexix. Plasma broadening is
then carried out for for each temperature and density of interest
throughout the solar and stellar interiors or HED plasma sources.

\section{Energy Dependence}

\begin{figure}
\includegraphics[height=3.0in,width=3.0in]{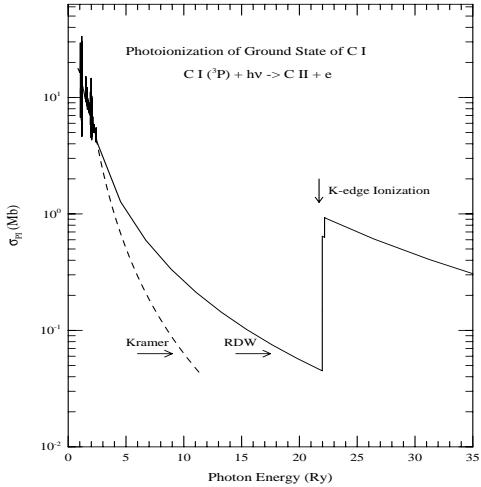}
\caption[Photoionization of \ci]{Photoionization cross section $\sigma_{PI}$ of the ground
state of C~I, $1s^22s^22p^2 \ ^3P$, computed using the relativistic
distorted wave (RDW) code by H.L. Zhang (discussed in \cite{aas})
compared with the Kramer's hydrogenic
formula Eq.~\ref{eq:kramer}. The large jump is due to
photoionization of the inner $1s$-shell or the K-edge. The resonance
structures at very low energies are obtained from the coupled channel
RM calculations in the Opacity Project.
\label{fig:pxci}}
\end{figure}

Photoionization cross sections vary widely in different approximations 
used to calculate opacities.
Simple methods such as the {\it quantum 
defect method} and the central-field approximation,
yield a feature-less background cross section. 
High-$n$ levels in a Rydberg series of levels 
behave hydrogenically at sufficiently high energies, and
the photoionization cross section may be approximated using
Kramer's formula (discussed in \cite{aas})
\begin{equation}
\sigma_{PI}= (\frac{8\pi}{3^{1.5}c}) \frac{1}{n^5\omega^3}.
\label{eq:kramer}
\end{equation}
Eq.~\ref{eq:kramer} is used in OP work to extrapolate photoionization cross 
sections in the high-energy region. However, it is not accurate, as seen
in Fig.~\ref{fig:pxci}. 
At high energies inner shells and 
sub-shells are ionized, and their contribution 
must also be included in total photoionization cross
sections. At inner (sub-)shell ionization thresholds
there is a sharp upward jump or edge and 
enhancement of the photoionization cross section. Fig.~\ref{fig:pxci}
shows results from a relativistic distorted wave (RDW) 
calculation and Kramer's fomula 
Eq.~\ref{eq:kramer}. The RDW results do not include resonances,
and differ from the OP results with resonance structures 
in the relatively small energy region
near the ioniization threshold.

\section{From Convergence to Completeness}

The NP16 work \cite{np16} also addressed an important point that
a reasonably complete expansion of
target configurations and levels in BPRM photoionization calculations
is necessary to ensure converged bound-free opacities.
The criteria for accuracy and completeness
are: (i) {\it convergence} of the wavefunction expansion (Eq.~6), 
and (ii) {\it  completeness} of PI cross sections, 
monochromatic and mean opacities with respect to
possibly large number of  multiply excited configurations.

 While NP16 demonstrated convergence
with respect to $n$=2,3,4 levels of the \fexviii target ion 
included in the RM 
calculations, more highly excited configurations that might affect 
high-energy behavior were not included. Subsequent work using and
comparing with the DW method was therefore carried out to ascertain the
effect of high-$n\ell$ configurations on opacities \cite{z18}.

Specifying excited configurations is straightforward in an atomic 
structure-DW calculation, but it is more complex and 
indirect in RM calculations. 
For example,                             
in order to investigate the role of more excited configurations the
NP16 BPRM calculations that yield 454 bound levels
\fexvii, were complemented with $>50\,000$ high$n,\ell$
"topup" levels      
to compute opacities and RMOs.
Photoionization cross sections of the 454 strictly bound levels computed
(negative 
eigenenergies) take into account embedded autoionizing resonances that are 
treated as distinct levels in DW calculations; therefore, in total there are 
commensurate number of levels to ensure completeness. 

 However, the large number of highly-excited 
configurations made only a small 
contribution to opacities, relative to the main BPRM cross 
sections, and only 
to the background cross sections.
without resonances. Therefore, the simpler
DW method may be used for topup cotributions without loss of accuracy as
to supplement RM calculations. Recent work has shown
that the topup contribution to RM opacities does not exceed 5\% to RMOs
\cite{petal23}.

\section{Sum Rule and Oscillator Strength Distribution}

The total $bb$ and integrated $bf$ oscillator strength, when
summed over all possible bb and bf transitions, must satisfy
the definition of the oscillator strength as fractional excitation 
probability, i.e. $\sum_j f_{ij} = N$, where $N$ is the number of active 
electrons.  But while the $f$-sum rule ensures completeness, 
it does not ensure accuracy of atomic calculations {\it per se}.  
That depends on the 
precise energy distribution of differential oscillator $df/dE$,
strength or photoionization cross section $\sigma_{PI}$. 
To wit: the hydrogenic approximation, if used for
complex atoms would satisfy the $f$-sum rule but would obviosuly be 
inaccurate. As disussed herein, the RM method is concerned primarily 
with $df/dE$ in the $bf$-continuum based on
full delineation of autoionizing resonance profiles.

As the end result, the RMO depends on energy distribution of
monochromatic opacity, convolved 
over the Planck function at a given temperature. 
Compared with OP results, the distribution of
RM \fexvii monochromatic opacity is quite different, and much more 
smoothed out without sharp variations that stem mainly from the treatment 
of resonances as $bb$ lines, even with limited autoionization 
broadening included perturbatively in DW opacity models. 
Experimentally, a flatter opacity distribution 
is also observed, in contrast to theoretical opacity models 
that exhibit larger dips in opacity at ``opacity windows'' 
\cite{b15,n19,np16,p18}.

\section{Conclusion}

This review describes photoionization work related to opacities.
The state-of-the-art R-matrix calculations are discussed in comparison
with the distorted wave data currently employed in opacity models.
Atomic and plasma effects such as channel coupling, broadening of
autoionizing resonances, high-energy behavior, and oscillator strength
sum-rule are described.

Existing OP and IP radiative data for
photoionization and transition probabilities for
astrophysically abundant elements have been
archived in databases TOPbase and TIPbase. OP opacities and radiative
accelerations are available online from OPserver \cite{topbase}.
R-matrix data for nearly 100 atoms and ions from uptodate and more
accurate calculations are available from the
database NORAD at OSU \cite{norad}.

\vskip 0.25in
{\bf Acknowledgements}
\vskip 0.25in

 I would like to thank Sultana Nahar for \fexvii atomic data and
discussions.


\vskip 0.25in
{\bf References}
\vskip 0.25in

\begin{thebibliography}{10}

\def\aa{{\it Astron. Astrophys.}\ }
\def\aasup{{\it Astron. Astrophys. Suppl. Ser.}\ }
\def\adndt{{\it Atom. data and Nucl. Data Tables.}\ }
\def\aj{{\it Astron. J.}\ }
\def\apj{{\it Astrophys. J.}\ }
\def\apjs{{\it Astrophys. J. Supp. Ser.}\ }
\def\apjl{{\it Astrophys. J. Lett.}\ }
\def\baas{{\it Bull. Amer. Astron. Soc.}\ }
\def\cpc{{\it Comput. Phys. Commun.}\ }
\def\jpb{{\it J. Phys. B}\ }
\def\jqsrt{{\it J. Quant. Spectrosc. Radiat. Transfer}\ }
\def\mn{{\it Mon. Not. R. astr. Soc.}\ }
\def\pasp{{\it Pub. Astron. Soc. Pacific}\ }
\def\pra{{\it Phys. Rev. A}\ }
\def\pr{{\it Phys.  Rev.}\ }
\def\prl{{\it Phys. Rev. Lett.}\ }
\def\hed{{\it High Energy Physics}\ }
\bibitem{m18} Mendoza, C.,
 Atoms {\bf 2018}, 6, 28.
\bibitem{pain23} Pain, J-C and Croset P., 
Atoms {\bf 2023}, 11, 27.
\bibitem{p18} Pradhan,A.K., 
 {\bf 2018}, ASP Conf. Ser. 515, 79-88
\bibitem{symp94} Seaton, M. J., Yu, Y., Mihalas,
D., \& Pradhan, A. K., {\bf 1994}, \mn 266, 805

\bibitem{op} {\em The Opacity Project}, The Opacity Project Team,
Institute of Physics Publishing, Vol. 1 (1995), Vol. 1 (1995)

\bibitem{b11} P.G. Burke, {\bf 2011},
{\it R-Matrix Theory of Atomic Collisions}, Springer
Series on Atomic, Optical and Plasma Physics.

\bibitem{aas} Pradhan A.K., Nahar S.N. {\em Atomic Astrophysics and
Spectroscopy}, {\bf 2011}, Cambridge University Press.

\bibitem{ip} Hummer DG, Berrington KA, Eissner W, Pradhan AK, Saraph HE,
Tully JA, {\em Astron. Astrophys.} {\bf 1993}; {\em 279}, 298-309
\bibitem{topbase} TOPbase, TIPbase:
Mendoza C, Seaton MJ, Buerger P, Bellorin P, Melendez M,
Gonzalez J, Rodriguez LS, Palacios E, Pradhan AK, Zeippen CJ,
{\bf 2007}, \mn, 378, 1031 
(http://cdsweb.u-strasbg.fr/topbase/topbase.html 
\bibitem{norad} Nahar SN, NORAD: Nahar-OSU-Radiative-Atomic-Data,
http://norad.astronomy.osu.edu, The Ohio State University
\bibitem{rm1} Berrington, K.A., Eissner, W., Norrington, P.H. .RMATRIX1:
Belfast atomic R-matrix codes. {\em Comput. Phys. Commun.} {\bf 1995}
{\em 92},
 290-420
\bibitem{np16} Nahar S.N., Pradhan A.K., 
\prl {\bf 2016} {\em 116}, 249502-294507.  
\bibitem{n11} Nahar S.N., Pradhan, A.K., Chen, G.-X and Eissner, W.
\pra {\bf 2011}, 83, 053417.
\bibitem{d21} Delahaye FD, Ballance CP, Smyth RT and Badnell NR,
\mn, {\bf 2021}, {\em 508}, 421-432. 
\bibitem{z18} Zhao L, Eissner W, Nahar SN, Pradhan AK, 
{\bf 2018}, ASP Conf. Ser. 515, 89-92.
\bibitem{s18} Nahar SN, {\bf 2018}, ASP Conf. Ser. 515, 93-103
\bibitem{ejn} Eissner W, Jones M, Nussbaumer H. {\bf 1974},
\cpc 8, 270-306
\bibitem{h75} Hibbert, A, {\bf 1974}, \cpc 9 141
\bibitem{p23} Pradhan  AK, {\bf 2023},
Astro-ph:https://arxiv.org/pdf/2301.07734.pdf
\bibitem{petal23} Pradhan  AK, Nahar SN, Eissner W, Zhao L (in
preparation).
\bibitem{b15} Bailey, J \etal, {\bf 2015}, Nature, 517,
56-59
\bibitem{n19} Nagayama, T \etal, {\bf 2019},
\prl, 122, 235001-1-6
\end{thebibliography}
\end{document}